\documentstyle{elsart}
\input epsf
\begin{document}
\begin{frontmatter}
\title{Volatility in Financial Markets:\\ 
Stochastic Models and Empirical Results}

\author{Salvatore Miccich\`e}, \author{Giovanni Bonanno}, 
\author{Fabrizio Lillo} and \author{Rosario N. Mantegna \thanksref{mail1}}

\address{
Istituto Nazionale per la Fisica della Materia, Unit\`a di Palermo\\
and\\
Dipartimento di Fisica e Tecnologie Relative,
Universit\`a di Palermo,\\
Viale delle Scienze, I-90128,
Palermo, Italia}



\thanks[mail1]{corresponding author, e-mail address: mantegna@unipa.it}

\begin{abstract}
We investigate the historical volatility of the 100 most capitalized
stocks traded in US equity markets. An empirical 
probability density function (pdf) of volatility is obtained and 
compared with the theoretical predictions of a lognormal model
and of the Hull and White model. The lognormal model well
describes the pdf in the region of low values of volatility 
whereas the Hull and White model better approximates the 
empirical pdf for large values of volatility. Both models
fails in describing the empirical pdf over a moderately 
large volatility range.      
\end{abstract}


\end{frontmatter}

\section{Introduction}

\indent Volatility of financial time series is a key 
variable in the modeling of financial markets. It controls 
all the risk measures associated with the dynamics of 
price of a financial asset. It also affects the 
rational price of derivative products. 
In this paper we consider some stochastic volatility models 
proposed in the financial literature by investigating their ability
in modeling statistical properties detected in empirical data.
Specifically, we investigate the probability density function (pdf)
of historical volatility for 100 highly capitalized stocks 
traded in the US equity markets. 
We observe that widespread volatility models such as the Hull and
White model \cite{HW1987} and the lognormal model fail in 
describing the volatility pdf when we ask the model to describe 
both low and high values of volatility. 
Our results show that a lognormal pdf better describes low values
of volatility whereas the Hull and White pdf gives a better approximation
of the empirical pdf for large values. 

\section{Volatility models}

The volatility $\sigma$ of a financial asset 
is a statistical quantity which 
needs to be determined
starting from market information \cite{Hull}.
It is the standard deviation
of asset return (or, almost equivalently, of logarithm
price changes of the asset). Different methodologies 
are used to infer
volatility estimation from market data ranging from a direct calculation
from past return data (historical volatility) to the computation
of the volatility implied in the determination of an option price computed
using the Black and Scholes formula \cite{BS1973} or some variant
of it.
There is a large empirical evidence that volatility 
is itself a stochastic process. 
In the present study we aim at comparing the theoretical predictions
for the pdf of the volatility $\sigma$ obtained with two different
stochastic volatility models with empirical 
observations obtained for the 100 most capitalized stocks 
traded in US equity markets (mostly the New York Stock
Exchange and the NASDAQ). 
The first model we consider is 
the Hull and White model \cite{HW1987}. In this model,
the variance rate $v \equiv \sigma^2$ is described 
by the Ito's equation
\begin{equation}
dv=a~(b-v)~dt+\xi v^{\alpha} dz_v
\end{equation}  
where $a$ and $b$ are parameters controlling the mean
reverting nature of the stochastic process, $\xi$ 
is controlling its diffusive aspects and $z_v$ is a Wiener 
process. The stochastic process is reverting at a level
$b$ at a rate $a$. The exponent $\alpha$ has been set to 1 in the
investigation of Hull and White \cite{HW1987}
and to 1/2 in the investigation of Heston \cite{Heston1993}.
In the present study we investigate the Hull and White
model with $\alpha=1$. The Ito's equation of this model for
the volatility $\sigma$ is
\begin{equation}
d\sigma=\frac{1}{2\sigma}\left\{\left[a~(b-\sigma^2)
-\frac{1}{2}\xi^2\sigma^2\right]~dt+
\xi \sigma^2 dz_{\sigma}\right\}
\end{equation}
This equation has been obtained starting from Eq. (1) 
and using Ito's lemma. The Hull and White model
has associated a stationary pdf of the volatility 
which has the form
\begin{equation}
P(\sigma)= 2 \frac{\left(ba/\xi^2\right)^{1+a/\xi^2}}{\Gamma(1+a/\xi^2)}
\frac{\exp(-ba/\xi^2 \sigma^2)}{\sigma^{2a/\xi^2+3}}
\end{equation} 
This pdf has a power-law tail for large values of $\sigma$. A power-law 
tail in the empirical volatility pdf has been observed in Ref. \cite{Liu99}
for large values of the volatility.
Another model is the lognormal model of 
volatility \cite{Cizeau97,Pasquini99}. 
An Ito's stochastic differential equation
associated with a lognormal pdf is
\begin{equation}
d\sigma=a~(b-\ln \sigma)~dt+\xi \sigma^{1/2} dz_{\sigma}
\end{equation} 
where $a$, $b$ and $\xi$ are control parameters 
of the model. 
The two models are characterized by quite different pdfs
especially for large values of the volatility where the 
Hull and White pdf shows a power-law behavior.
The present study aims at detecting the regions of validity 
of these two models in empirical data.
\begin{figure}[t]
\epsfxsize=5.0in
\centerline{\epsfbox{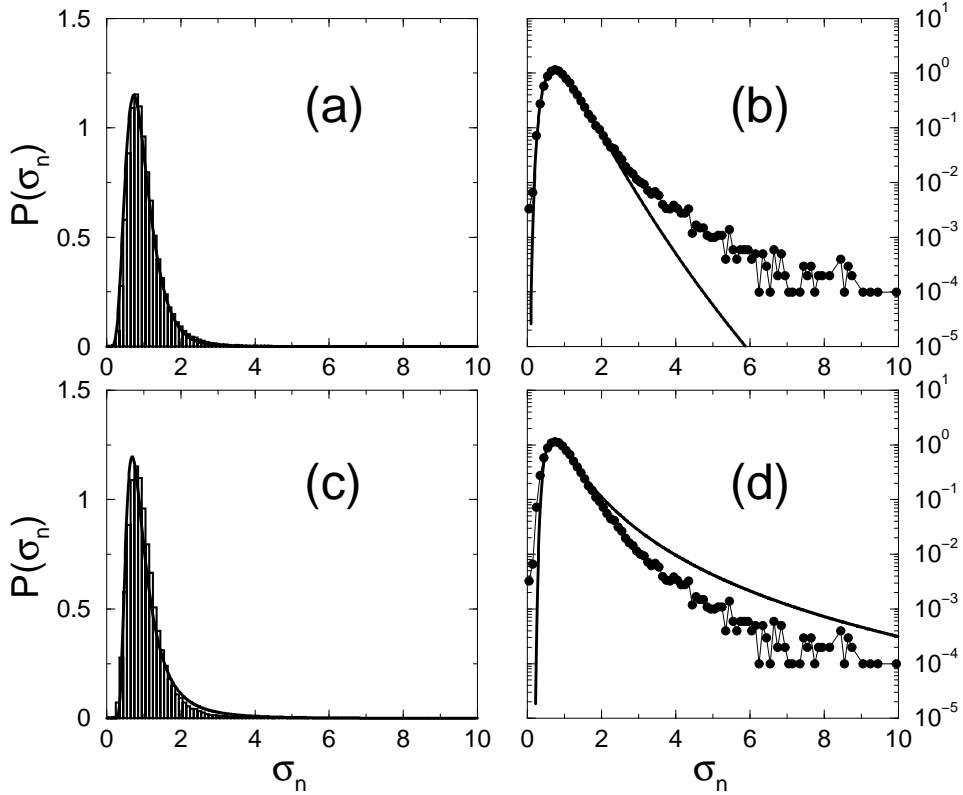}}
\caption{Best fits of the empirical pdf of normalized volatility 
obtained by investigating 100 stocks traded in US equity markets
during the time period
January 1995 - December 1998. In panel (a) and (c) the y axis is linear whereas in panel 
in (b) and (d) the data are shown in a semilogarithmic plot. 
In all panels, the solid lines are the best fits whereas the histogram and solid 
circles are empirical data. 
In panels (a) and (b) we show the best
fittings obtained with a lognormal pdf of mean value 0.97  and variance
equals to 0.19 and in panels (c) and (d) we show the best fittings obtained
with the Hull and White pdf of Eq. (3). In this last case 
the fittings parameters are $2a/\xi^2+3=3.79$ and $ba/\xi^2=0.91$.}
\label{fig1}
\end{figure}
\section{Empirical Results}
With this goal in mind, we investigate the statistical properties 
of volatility for the 100 most capitalized stocks traded in 
US equity markets during 
a 4 year time period. The empirical data are taken from 
the trade and quote (TAQ) database, maintained by
the New York Stock Exchange (NYSE). In particular, 
our data cover the whole period ranging from January $1995$ to 
December $1998$ ($1011$ trading days). This database contains 
all transactions occurred for each stock traded
in the US equity markets.
The capitalization considered is the one recorded on 
August 31, $1998$. For each stock and for each trading day 
we consider the time series of stock price recorded 
transaction by transaction. Since transactions for different stocks
do not happen simultaneously, 
we divide each
trading day (lasting $6^h~30'$) into $12$ intervals of $1950$ 
seconds each. In
correspondence to each interval, we define $12$ (intraday)
stock's prices proxies $S(k)$ -- with $k=1, \cdots, 12$ defined  
as the transaction price detected nearest to end of the interval
(this a one possible way to deal with high-frequency financial
data \cite{Dacorogna2001}). 
\begin{figure}[t]
\epsfxsize=4.8in
\centerline{\epsfbox{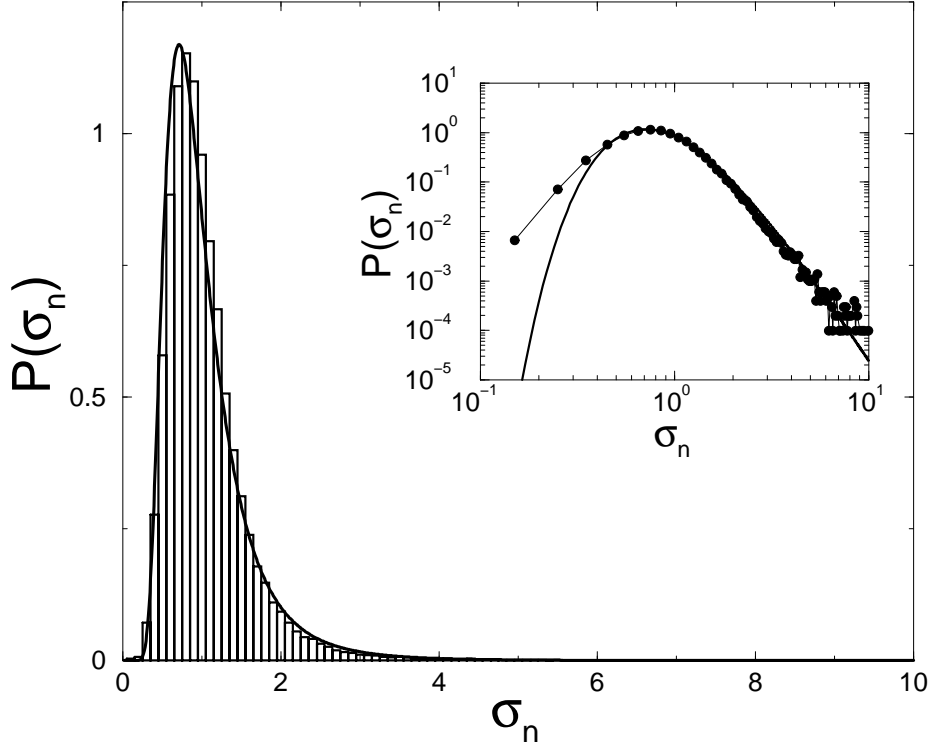}}
\caption{Empirical pdf of normalized volatility compared with the pdf
predicted by the stochastic model of Eq. (5). In the figure 
the y axis is linear whereas in the inset the data are shown 
in a log-log plot. 
Solid lines represent the best fit whereas the histogram and solid 
circles are empirical data. The 
fittings parameters of Eq. (6) are $a/\xi^2+2=6.27$ 
and $ba/\xi^2=4.45$.}
\label{fig2}
\end{figure}
We choose 12 intraday intervals since this value ensures that  
at least one transaction is in average observed in each interval
for all considered stocks in the present study. 
For each stock we can thus compute a historical 
daily volatility as 
$\sigma(t) = \sqrt{11}~{\rm std}[\ln (S(k+1)/S(k))]$,
where ${\rm std}[\cdot]$ indicates the standard deviation
of the argument of the function.
Hence, for each stock we have $1011$ values of daily volatility. 
These volatility data have then been analyzed to compute the 
volatility pdf for each stock. The 100
empirical pdfs we obtain are then fitted with the theoretical pdfs
of the considered models. 
Due to the limited number of records used to estimate the empirical 
pdfs (1011 records per stock) the results of our fittings are not 
able to indicate strengths and weaknesses of the two models.
For this reason we rescale the volatility value of each stock $\sigma$
to its mean value $<\sigma>$ and we investigate the pdf of the
normalized variable $\sigma_n=\sigma/<\sigma>$ for the ensemble of 100 
stocks. In this way we obtain an empirical pdf which is quite 
accurate being based on the recording of 101,100 events.
The best fittings of this empirical
pdf with Eq. (3) and with a lognormal pdf 
are shown in Fig. 1. It should be noted that 
the form of Eqs (2) and (4) implies that the three control parameters
of these equations reduce to two independent fitting parameters.  
For the lognormal pdf 
the fitting parameters can be chosen as
the mean and the variance whereas for the Hull and White model the fitting
parameters can be written as $2a/\xi^2+3$ and $ba/\xi^2$. 
The top two panels show the best
fittings of the lognormal pdf. The lognormal pdf describes very well
low values of volatility in the interval $0<\sigma_n\le 2$ but
completely fails in describing large values $\sigma_n > 2$.
In particular a lognormal pdf underestimates large values of volatility.
The bottom two panels show the best fits obtained with the
Hull and White pdf. In this case the volatility low values are only
approximately well described by the theoretical pdf. Moreover, for large values
of volatility, the best fit overestimates by approximately a factor two
empirical results. 
Both the lognormal model and the Hull and
White models fail in describing well the normalized volatility over a
relatively wide range of volatility values ($0<\sigma_n<10$).
This implies that there is still room for the improvement of 
volatility models down to the basic aspect of well describing 
the asymptotic pdf of volatility over a realistically wide
range. For example, we have verified that a volatility model 
described by the Ito's equation
\begin{equation}
d\sigma=a~(b-\sigma)~dt+\xi \sigma dz_v
\end{equation} 
is characterized by a pdf which has intermediate properties 
to the ones of the pdfs of the two volatility models 
investigated in this paper. This model predicts a pdf of 
the volatility which has the form
\begin{equation}
P(\sigma)= \frac{\left(ba/\xi^2\right)^{1+a/\xi^2}}{\Gamma(1+a/\xi^2)}
\frac{\exp(-ba/\xi^2 \sigma)}{\sigma^{a/\xi^2+2}}
\end{equation} 
This pdf has also a power-law tail for large values of $\sigma$
but it predicts a different shape for low values of $\sigma$ .
The two independent fitting parameters of this pdf
can be written in terms of $a/\xi^2$ and $ba/\xi^2$.
In Fig. 2 we show the best fit
with the pdf of Eq. (6). The agreement
with empirical data is rather good in a volatility range from $\sigma_n=0.5$
to $\sigma_n=10$. 

In summary, we report on a comparison of two widespread 
theoretical 
models of volatility with empirical data obtained by collecting 
together the volatility of 100 most capitalized stocks traded in 
US equity markets. The comparison is focused on the shape of the asymptotic
pdf of volatility. Two widespread models (lognormal and Hull and White)
fail in describing the 
pdf over a relatively wide volatility range. We show that
the model of Eq. (5)  
improves the overall description of the pdf especially for values 
of normalized volatility $\sigma_n>0.5$. 
Further research 
attempts are needed to select the most appropriate Ito's 
model able to describe volatility both under the 
aspects of the pdf and under the dynamics aspects of 
the nature and form of volatility auto-correlation function.
Indeed, there is a growing evidence that the volatility autocorrelation
function is long-range correlated \cite{Dacorogna2001}
and this key aspect is not taken
into account in most of the widespread models of volatility 
(as the ones considered 
in the present study) which are typically characterized 
only by short-range time memory.
\begin{ack}
The authors thank INFM, MIUR and ASI for financial support. This work 
is part of the FRA-INFM project 'Volatility in financial markets'. 
\end{ack}

\end{document}